\documentclass[%
 aip,
 amsmath,amssymb,
 reprint,
 twocolumn,
 dblfloatfix]{revtex4-2}

\usepackage{graphicx}
\usepackage{dcolumn}
\usepackage{bm}
\usepackage{siunitx}
\usepackage{amsmath}

\usepackage[usenames,dvipsnames,svgnames,table]{xcolor}
\colorlet{m}{black}
\colorlet{k}{Black}

\usepackage[textwidth=12.5mm]{todonotes}
\usepackage{easyReview}
\usepackage[colorlinks=true,citecolor=blue]{hyperref}
\usepackage[rightcaption]{sidecap}

\begin{document}

\preprint{AIP/123-QED}

\title{Generalized quantum limits of electrical contact resistance and thermal boundary resistance}

\author{Alice Ho}
\affiliation{\hbox{School of Electrical and Computer Engineering, Cornell University, Ithaca, NY, 14853, USA}}
\author{Jashan Singhal}
\affiliation{\hbox{School of Electrical and Computer Engineering, Cornell University, Ithaca, NY, 14853, USA}}
\affiliation{\hbox{Now at Intel Corporation}}
\author{Deji Akinwande}
\affiliation{\hbox{Department of Electrical and Computer Engineering, University of Texas at Austin, Austin, TX 78712, USA}}
\author{Huili G. Xing}
\affiliation{\hbox{School of Electrical and Computer Engineering, Cornell University, Ithaca, NY, 14853, USA}}
\affiliation{\hbox{Department of Materials Science and Engineering, Cornell University, Ithaca, NY, 14853, USA}}
\affiliation{\hbox{Kavli Institute at Cornell for Nanoscale Science, Cornell University, Ithaca, NY, 14853, USA}}
\author{Debdeep Jena}
\affiliation{\hbox{School of Electrical and Computer Engineering, Cornell University, Ithaca, NY, 14853, USA}}
\affiliation{\hbox{Department of Materials Science and Engineering, Cornell University, Ithaca, NY, 14853, USA}}
\affiliation{\hbox{Kavli Institute at Cornell for Nanoscale Science, Cornell University, Ithaca, NY, 14853, USA}}

\begin{abstract}
The importance of electrical contact resistance and thermal boundary resistance has increased dramatically as devices are scaled to atomic limits.  The use of a rich range of materials with various bandstructures (e.g. parabolic, conical), and in geometries exploiting various dimensionalities (e.g. 1D wires, 2D sheets, and 3D bulk) will increase in the future.  Here we derive a single general expression for the quantum limit of electrical contact resistivity for various bandstructures and all dimensions.  A corresponding result for the quantum limit of thermal boundary resistance is also derived.  These results will be useful to quantitatively co-design, benchmark, and guide the lowering of electrical and thermal boundary resistances for energy efficient devices.
\end{abstract}

\maketitle

\begin{figure*}
	\includegraphics[width=0.9\textwidth]{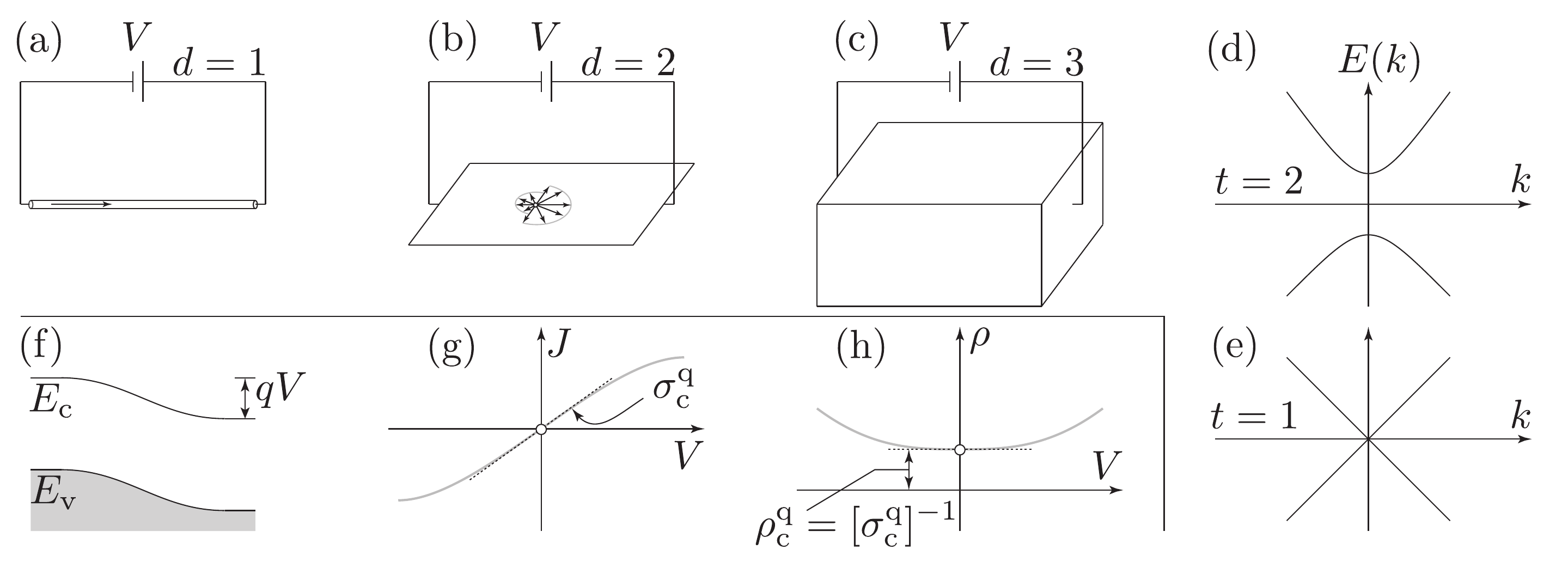} 
	\caption{\label{fig1} Schematic representation of (a) 1D, (b) 2D, and (c) 3D channels with (d) parabolic with $t=2$ and (e) linear with $t=1$ energy bandstructure which is written as $E(k) = [ \sum_{i=1}^{d} (\alpha_i k_i)^2 ]^{\frac{t}{2}}$, where $\alpha_i = \hbar v_{{\rm F}i}$ for linear and $\alpha_{i} = \hbar / \sqrt{2 m_{i}}$ for parabolic bands.  The schematic energy band diagram in (f) indicates the current response $J-V$ curve of (g), and results in the contact resistivity in (h).
	}
\end{figure*}

In studies of semiconductors with parabolic bandstructure Maasen {\em et al.}\cite{maassen2013full} and Barasker {\em et al.}\cite{baraskar2013lower} derived separate expressions for the quantum limits of electrical contact resistivity for $d=$1, 2, and 3 dimensions. Recently Singhal and Jena (SJ) derived a {\em single} unified expression for particle, charge, and heat currents in the quantum limit for various bandstructures and dimensions\cite{singhal2020unified}.  Using \cite{singhal2020unified}, in this letter we derive a {\em single} unified expression for the quantum limit of electrical contact resistivity for any dimension ($d=1,2,3$) and for two bandstructures: parabolic $E(k) = \hbar^2 k^2 / 2 m^{\star}$ and conical $E(k) = \hbar v_{\rm F} k$ as indicated in Figures \ref{fig1}(a)-(e).  The general results derived in this work apply to anisotropic bands of each type in all dimensions and temperatures.  A general quantum limit of thermal boundary resistance is also derived using\cite{singhal2020unified}.  The results address the increased need of co-design of electro-thermal phenomena in devices in their atomic limits both at room temperature and in the cryogenic limit.

The unified SJ ballistic equation for particle, charge, and energy currents $J=g\sum_{k}v(k)^{a}E(k)^{b}f(k)$, a sum over states $k$ of group velocity $v(k)$, energy $E(k)$, and occupation $f(k)$ in the ballistic limit, is derived in \cite{singhal2020unified} to be 
\begin{equation}
    J_{d,t}^{a,b} = \frac{g}{\lambda^{d} \beta^b} \cdot (\frac{\lambda_{1}}{h \beta})^{a} \cdot C_{d,t}^{a,b} \cdot F_{j}^{\pm}(\eta),
    \label{general_current}
\end{equation}
where $\lambda$ is the de-Broglie wavelength of electrons (or phonons), $h$ is Planck's constant, $\beta=1/(k_{\rm b}T)$, and $F^{\pm}_{j}(\eta) = \frac{1}{\Gamma(j+1)}  \int_{0}^{\infty}\,\mathrm{d}x \frac{x^{j}}{ \exp{[x - \eta]} \pm 1 }$ is the Fermi-Dirac (+) or Bose-Einstein (-) integral. 
All terms and units of Equation \ref{general_current} are explicitly defined in \cite{singhal2020unified} for parabolic ($t=2$) and conical ($t=1$) energy dispersions.  

For charge current ($a=1$, $b=0$) of {\em fermions}, Eq. \ref{general_current} is
\begin{equation}
J_{d,t}^{1,0}(v) = g \cdot \frac{q}{h} \cdot \frac{\lambda_1}{\lambda^d} \cdot \frac{\Gamma(r+1)}{\Gamma(\frac{d+1}{2})} \cdot [ \frac{ F^{+}_{r}(\eta) - F^{+}_{r}(\eta - v) }{\beta} ], 
\end{equation}
where $r=(d-1)/t$, $q$ the electron charge, and $g=g_{\rm s} g_{\rm v}$ is the product of the spin degeneracy $g_{\rm s}$ and the valley degeneracy $g_{\rm v}$. $\lambda_1$ is the de-Broglie wavelength in the direction of net current flow\cite{de_broglie_note} and $\Gamma(...)$ is the gamma function. $\eta = \beta E_{\rm F}$ is the dimensionless chemical potential and $v=\beta qV$ is the dimensionless voltage for which the fermionic current flows in response.  The quantum contact conductivity  $\sigma_{\rm c}^{\rm q} = \lim_{V \rightarrow 0} J_{d,t}(v)/V$  
\begin{eqnarray}
\sigma_{\rm c}^{\rm q} = g \cdot \frac{q^2}{h} \cdot \frac{\lambda_1}{\lambda^d} \cdot \frac{\Gamma(r+1)}{\Gamma(\frac{d+1}{2})} \cdot \underbrace{ \lim_{v \rightarrow 0}  [ \frac{ F^{+}_{r}(\eta) - F^{+}_{r}(\eta - v) }{v} ] }_{ \partial F_{r}^{+}(\eta)/\partial \eta = F_{r-1}^{+}(\eta) } \nonumber\\
\implies \sigma_{\rm c}^{\rm q} = g \cdot \frac{q^2}{h} \cdot \frac{\lambda_1}{\lambda^d} \cdot \frac{\Gamma(\frac{d-1}{t}+1)}{\Gamma(\frac{d+1}{2})} \cdot F_{\frac{d-1}{t}-1}^{+}(\eta), \text{ }
\label{contact_gen}
\end{eqnarray}
and the mobile fermion density
\begin{equation}
n_{d,t} = 2 J_{d,t}^{0,0} = \frac{2g}{\lambda^d} \cdot  \frac{\Gamma(\frac{d}{t})}{t \Gamma(\frac{d}{2})} \cdot F_{\frac{d}{t}-1}^{+}(\eta)
\label{charge_gen}
\end{equation}
form a pair of {\em exact} closed-form analytical expressions of measurable quantities for all dimensions and temperatures.  Combining Equation \ref{contact_gen} with \ref{charge_gen} helps express the exact generalized quantum contact conductivity limit as 
\begin{equation}
\boxed{ \sigma_{\rm c}^{\rm q} = \big[g^{\frac{1}{d}} \cdot \frac{q^2}{h} \cdot \frac{\lambda_1}{\lambda} \cdot{(n_{d,t})^{\frac{d-1}{d}}}\big] \cdot H_{d,t}(\eta), }
\label{contact_gen_final}
\end{equation}
where the unitless factor $H_{d,t}(\eta)$ is 
\begin{equation}
H_{d,t}(\eta)= \frac{\Gamma(\frac{d-1}{t}+1)}{\Gamma(\frac{d+1}{2})} \cdot [\frac{\frac{t}{2} \Gamma(\frac{d}{2}) }{ \Gamma(\frac{d}{t}) } ]^{\frac{d-1}{d}} \cdot \frac{F_{\frac{d-1}{t}-1}^{+}(\eta)}{ [F_{\frac{d}{t}-1}^{+}(\eta)]^{\frac{d-1}{d}} }.
\label{ratio}
\end{equation}


Low-resistance contacts are made to degenerately doped materials for which $\eta >> +1$ and the Fermi-Dirac integrals are $F_{j}^{+}(\eta) \approx \eta^{j+1}/\Gamma(j+2)$. Then, $H_{d,t}(\eta \rightarrow \infty) = [\Gamma(\frac{d+2}{2})]^{(d-1)/d}/\Gamma(\frac{d+1}{2})$, independent of both $\eta$ and the bandstructure $t$, depending only on $d$. The degenerate limit of the quantum limit of contact resistance for isotropic bandstructure ($\lambda_1 = \lambda$) from Equation \ref{contact_gen_final} written explicitly then is
\begin{equation}
    \boxed{\rho_{\rm c}^{\rm q} = (\sigma_{\rm c}^{\rm q})^{-1} \approx \frac{h}{q^2} \cdot \frac{1}{(g_{\rm s} g_{\rm v})^{\frac{1}{d}}} \cdot \frac{\Gamma(\frac{d+1}{2})}{[\Gamma(\frac{d+2}{2})]^{\frac{d-1}{d}}} \cdot \frac{1}{(n_{\rm d})^{\frac{d-1}{d}}}, }
    \label{degenerate_limit_general}
\end{equation}
which is universal, and valid for both parabolic ($t=2$) and conical ($t=1$) bandstructures for all dimensions $d$. To a good approximation, $[\Gamma(\frac{d+2}{2})]^{(d-1)/d}/\Gamma(\frac{d+1}{2}) \approx d^{1/6}$ for $d=1,2,3$. This reveals that the generalized degenerate limit of the quantum limit of electrical contact conductivity is the rather simple relation
\begin{eqnarray}
\sigma_{\rm c}^{\rm q} \approx (g_{\rm s}g_{\rm v})^{\frac{1}{d}} \cdot \frac{q^2}{h} \cdot  d^{\frac{1}{6}}  \cdot (n_{d})^{\frac{d-1}{d}}.
\label{degenerate_limit}
\end{eqnarray}
The generalized result Equation \ref{contact_gen_final} of the electronic quantum contact conductance, and its degenerate limit relations Equations \ref{degenerate_limit_general} and \ref{degenerate_limit} are the main results of this work.  The ratio of gamma functions in Equation \ref{ratio} is unity for parabolic bandstructure ($t=2$), and the exact generalized quantum conductivity for this case is
\begin{equation}
\boxed{ \sigma_{\rm c}^{\rm q} = (g_{\rm s}g_{\rm v})^{\frac{1}{d}} \cdot \frac{q^2}{h} \cdot  \frac{\lambda_1}{\lambda} \cdot (n_{d})^{\frac{d-1}{d}} \cdot \frac{ F_{\frac{d-3}{2}}^{+}(\eta) }{ [F_{\frac{d}{2}-1}^{+}(\eta)]^{\frac{d-1}{d}} }. }
\end{equation}

Figure \ref{fig1} (b) schematically indicates the group velocity vectors of carriers. The right-moving states are in equilibrium with the left electrode whose chemical potential is maintained at an energy $qV$ higher by the voltage source than the right electrode, with which the left-moving states are in equilibrium.  Figure \ref{fig1} (f) indicates an energy band diagram and Figure \ref{fig1} (g) shows the current-voltage response whose $V\rightarrow 0$ slope is the quantum limit of conductivity $\sigma_{\rm c}^{\rm q}$ given by Equation \ref{contact_gen_final}. The $V\rightarrow 0$ limit of electrical resistivity $\rho = 1/(\partial J/\partial V)$ shown in Figure \ref{fig1} (h) is the quantum limit of the contact resistance, $\rho_{\rm c}^{\rm q} = 1/\sigma_{\rm c}^{\rm q}$.

\begin{figure*}
	\includegraphics[width=\textwidth]{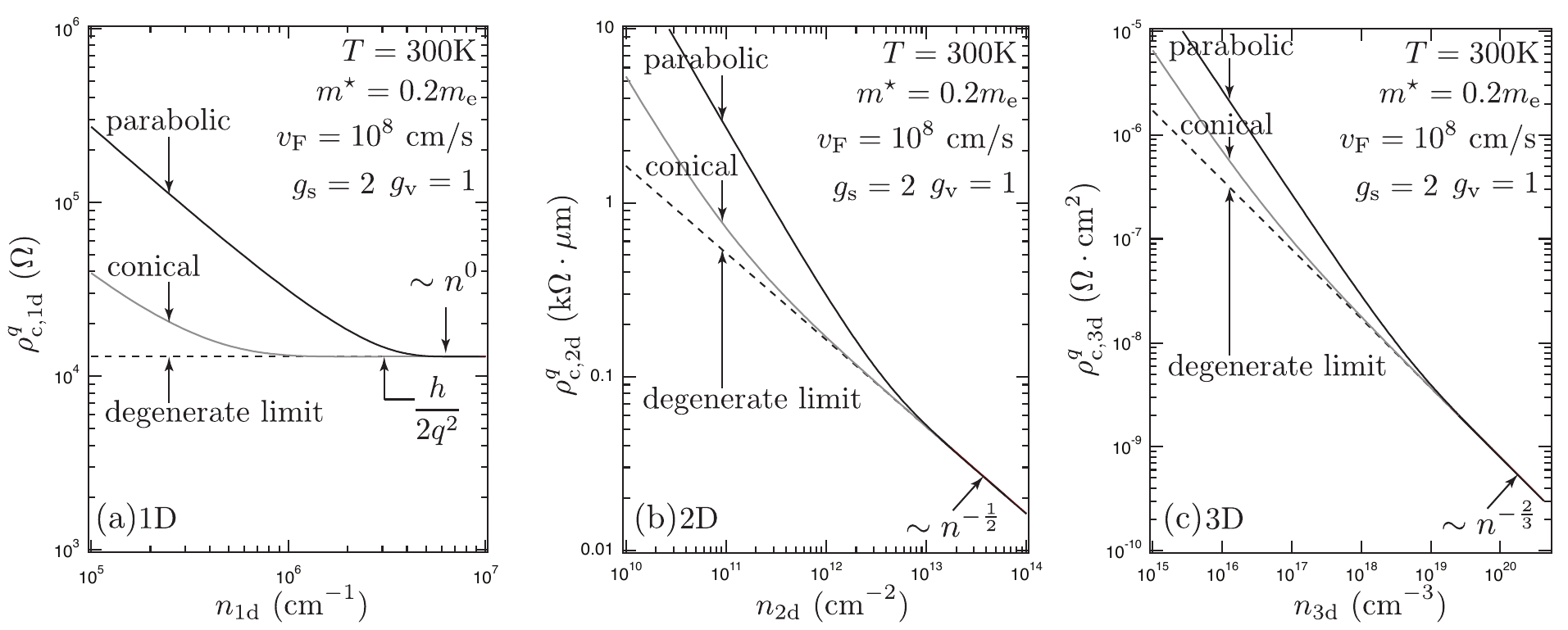} 
	\caption{\label{fig2} The quantum limit of electrical contact resistance for (a) 1D, (b) 2D, and (c) 3D conducting channels.  Each of the three plots compare values at $T=$ 300 K for parabolic bandstructure with $m^{\star}=0.2 m_{\rm e}$ shown as thick solid lines, conical bands with Fermi velocity $v_{\rm F}=10^{8}$ cm/s shown a thin solid gray lines, each with spin degeneracy $g_{\rm s}=2$ and valley degeneracy $g_{\rm v} = 1$.  The degenerate limit from Equation \ref{degenerate_limit} is shown as the dashed line.
	}
\end{figure*}

Equation \ref{contact_gen_final} indicates a transition from $\rho_{\rm c}^{\rm q} \propto n_{d}^{-1}$ for all $d$ for low $n_{d}$ to $\rho_{\rm c}^{\rm q} \propto n_{d}^{-(d-1)/d}$ for high $n_{d}$.  While Equations \ref{degenerate_limit_general} and \ref{degenerate_limit} are applicable in desired ultralow resistance ohmic contacts due to high carrier concentrations, for low carrier density contacts Equation \ref{contact_gen_final} must be used.  Figure \ref{fig2} shows the calculated quantum limits of contact resistance as a function of the carrier densities using Equation \ref{contact_gen_final} in solid lines, and its degenerate limit using Equation \ref{degenerate_limit} as dashed lines.  Figure \ref{fig2}(a) for a $d=1$ channel shows that in the degenerate limit, the quantum limit of the 1D contact resistance is the Landauer or Sharvin limit which also appears experimentally in quantized conductance, the integer quantum Hall effect, and the conductance per subband in metallic single-wall carbon nanotubes\cite{wong2010carbon}.  

Though Equation \ref{degenerate_limit} shows that for $d=1$, $\rho_{\rm c}^{\rm q} \rightarrow h/(gq^2)$ in the degenerate limit is independent of the 1D carrier density, from Equation \ref{contact_gen_final} and Figure \ref{fig2}(a), the quantum contact resistance limit increases inversely with the carrier density $\rho_{\rm c}^{\rm q} \propto 1/n_{\rm 1d}$ in the non-degenerate limit, for both parabolic and conical bandstructure.  Thus for a doubly degenerate spin 1D channel with a parabolic bandstructure with $m^{\star}=0.2 m_{\rm e}$ at 300 K, $\rho_{\rm c}^{\rm q}=h/(2q^2)=12.6$ k$\Omega$ for $n_{\rm 1d}>5 \times 10^{6}$/cm, which increases to $\rho_{\rm c}^{\rm q} \approx 30$ k$\Omega$ for $n_{\rm 1d}=  10^{6}$/cm, at which carrier density for the conical bandstructure with $v_{\rm F}=10^{8}$ cm/s, the contact resistance stays at $\rho_{\rm c}^{\rm q}=12.6$ k$\Omega$.  This comparison shows that care must be exercised in using the degenerate limit result of Equation \ref{degenerate_limit} shown as the dashed line in Figure \ref{fig2} (a), because it is valid only in the degenerate limit, whereas Equation \ref{contact_gen_final} is always valid.

The points made for the quantum limit of contact resistance for the 1D case in Figure \ref{fig2}(a) are reinforced in Figure \ref{fig2}(b) for 2D channels, and in Figure \ref{fig2}(c) for 3D channels.  The degenerate quantum limit of the contact resistance goes as $\sim n_{d}^{-(d-1)/d}$ according to Equation \ref{degenerate_limit_general}.  Therefore, while for $d=1$ this has no dependence on the 1D carrier density ($\rho_{\rm c}^{\rm q} \sim n_{\rm 1d}^{0}$), Figure \ref{fig2}(b) indicates for $d=2$ the dependence is $\rho_{\rm c}^{\rm q} \sim n_{\rm 2d}^{-1/2}$, and Figure \ref{fig2}(c) shows for $d=3$ it is $\rho_{\rm c}^{\rm q} \sim n_{\rm 3d}^{-2/3}$.  In the non-degenerate limit at low carrier concentrations, all quantum limited contact resistances follow a universal dependence on the carrier density of $\rho_{\rm c}^{\rm q} \sim n_{d}^{-1}$, both for parabolic and conical bandstructures, though the precise values may differ based on the bandstructure parameters $m^{\star}$ for parabolic and $v_{\rm F}$ for conical.  This dependence changes to the  universal dependence of $\sim n_{d}^{-(d-1)/d}$ in the degenerate limit, where the bandstructure parameters $m^{\star}$ for parabolic and $v_{\rm F}$ for conical do not matter anymore.

What is not explicitly shown in Figure \ref{fig2} is the dependence of $\rho_{\rm c}^{\rm q}$ on the spin and valley degeneracies $(g_{\rm s}, g_{\rm v})$ of the bandstructure, and anisotropies.  Equations \ref{contact_gen_final} and \ref{degenerate_limit_general} show that these properties do have an effect on the precise values of $\rho_{\rm c}^{\rm q}$ at all carrier densities, but do not change the asymptotic power law dependencies on the carrier densities at the two extremes.  Figure \ref{fig2} shows that for semiconductor channels in the degenerate limit, the lowest contact resistances for each valley with $g_{\rm s}=2$ are $\sim 12.6$ k$\Omega$ for 1D, $\sim 0.05$ k$\Omega \cdot\mu$m (= $\sim 0.05$ $\Omega \cdot$mm = 50 $\Omega \cdot\mu$m for $n_{\rm 2d} \sim 10^{13}$/cm$^{2}$), and $\sim 8 \times 10^{-10}$ $\Omega \cdot$cm$^{2}$ for $n_{\rm 3d}\sim 10^{20}$/cm$^{3}$.  

As emphasized in\cite{maassen2013full,baraskar2013lower,jena2014intimate,akinwande2025quantum}, because the quantum limit formulae here assumes unity transmission coefficient, the presence of tunneling barriers increase the actual $\rho_{\rm c}$ from the limits $\rho_{\rm c}^{\rm q}$ derived and plotted in Figure \ref{fig2}.  The conical bandstructure discussed here is not limited to graphene-like Dirac cone bandstructures because the bandstructure of narrow bandgap semiconductors far from the band edges approaches linear $E(k)$ dependence\cite{jena2022quantum}.  A detailed comparison of current state of the art experimental contact resistances for various materials of different dimensionalities, anisotropies, and bandstructures, will presented in a separate comprehensive article expanding the results discussed in this letter.


\begin{table*}
\caption{\label{tab:quantum_limits_electrical_contact_resistance_thermal_boundary_resistance}Quantum limits of electrical contact resistance and thermal boundary resistance in various dimensions}
\begin{ruledtabular}
\begin{tabular}{cccc}
      &  1D & 2D & 3D  \\ 
    \hline
                                    &   & &   \\
    $\sigma_{\rm c}^{\rm q}$ (parabolic, exact)   & $ \frac{q^2}{h}(g_{\rm s}g_{\rm v}) F_{-1}^{+}(\eta)$  & $\frac{q^2}{h}(g_{\rm s}g_{\rm v})^{\frac{1}{2}} \frac{\lambda_1}{\lambda} (n_{\rm 2d})^{\frac{1}{2}}  \frac{F_{-\frac{1}{2}}^{+}(\eta)}{[F_{0}^{+}(\eta)]^{\frac{1}{2}}}$ & $\frac{q^2}{h} (g_{\rm s}g_{\rm v})^{\frac{1}{3}} \frac{\lambda_1}{\lambda} (n_{\rm 3d})^{\frac{2}{3}} \frac{F_{0}^{+}(\eta)}{[F_{\frac{1}{2}}^{+}(\eta)]^{\frac{2}{3}}}$ \\
                            &   &  &   \\
    $\sigma_{\rm c}^{\rm q}$ (conical, exact)   & $\frac{q^2}{h} (g_{\rm s}g_{\rm v}) F_{-1}^{+}(\eta)$  & $\frac{q^2}{h}(g_{\rm s}g_{\rm v})^{\frac{1}{2}} \frac{\lambda_1}{\lambda} (n_{\rm 2d})^{\frac{1}{2}}  \sqrt{\frac{2}{\pi}} \frac{F_{0}^{+}(\eta)}{[F_{+1}^{+}(\eta)]^{\frac{1}{2}}}$ & $\frac{q^2}{h}(g_{\rm s}g_{\rm v})^{\frac{1}{3}} \frac{\lambda_1}{\lambda} (n_{\rm 3d})^{\frac{2}{3}} \frac{\pi^{\frac{1}{3}}}{4} \frac{F_{+1}^{+}(\eta)}{[F_{+2}^{+}(\eta)]^{\frac{2}{3}}}$ \\
                            &   &  &   \\
    $\rho_{\rm c}^{\rm q}$ (degenerate)   & $\frac{h}{q^2} \frac{1}{g_{\rm s}g_{\rm v}}$  & $ \frac{h}{q^2} \sqrt\frac{\pi}{4 g_{\rm s}g_{\rm v} n_{\rm 2d}}$ & $\frac{h}{q^2} (\frac{16}{9\pi g_{\rm s}g_{\rm v} (n_{\rm 3d})^2})^{\frac{1}{3}}$ \\
                            &   &  &   \\
    $ \kappa_{\rm c}^{\rm q} $   & $ \frac{\pi^2}{3} \frac{k_{\rm b}^2}{h} T $  & $ 12 \zeta(3) \frac{k_{\rm b}^3}{h^2 v} T^2 $ & $ \frac{4 \pi^5}{15} \frac{k_{\rm b}^4}{h^3 v^2} T^3 $ \\
                            &   &  &   \\
    $ \frac{\kappa_{\rm c}^{\rm q} \rho_{\rm c}^{\rm q}({\rm deg.})}{T^d} $   & $ \frac{\pi^2}{3g_{\rm s}g_{\rm v}} \frac{k_{\rm b}^2}{q^2} $  & $ 6 \zeta(3) \sqrt\frac{\pi}{ g_{\rm s}g_{\rm v} n_{\rm 2d}} \frac{k_{\rm b}^3  }{h v q^2} $ & $ \frac{4 \pi^5}{15} (\frac{16}{9\pi g_{\rm s}g_{\rm v} (n_{\rm 3d})^2})^{\frac{1}{3}} \frac{k_{\rm b}^4 }{h^2 v^2 q^2} $ \\
                            &   &  &   \\
\end{tabular}
\end{ruledtabular}
\end{table*}

To obtain the generalized quantum limit of the thermal boundary resistance\cite{swartz1989thermal,chen2022interfacial,cahill2003nanoscale}, the heat current requires higher moment terms with $(a,b)=(1,1)$ in the SJ ballistic Equation \ref{general_current}.  The general expression in $d$-dimensions for the ballistic thermal boundary resistance is obtained from the heat current $Q_{d, t} = J_{d,t}^{1,1} - \mu J_{d,t}^{1,0}$ flowing from an electrode at chemical potential $\mu$ and temperature $T$.  The heat current is carried primarily by phonons (=bosons) in a semiconductor. Then the chemical potential $\mu = 0$ and the low-energy dispersion $\omega(k) = v k$ for each acoustic phonon branch with $t=1$ is characterized by a sound velocity $v$ of that branch.  

The heat current density per acoustic phonon branch is then
\begin{equation}
    Q_{d} = \frac{g \pi^{\frac{d-1}{2}} k_{\rm b}^{d+1} }{ h^d v^{d-1} } \cdot \frac{\Gamma(d+1)}{\Gamma(\frac{d+1}{2})} \cdot F_{d}^{-}(0) \cdot T^{d+1},
\end{equation}
where the Bose-Einstein integral $F_{d}^{-}(...)$ is related to the Riemann zeta function $\zeta(...)$ via $F_{d}^{-}(0)=\zeta(d+1)$.  The quantum limit of the thermal boundary conductance $\kappa_{\rm c}^{\rm q} = \partial Q_{d}  / \partial T$ in $d-$dimensions for each phonon branch then becomes 
\begin{eqnarray}
    \boxed{\kappa_{\rm c}^{\rm q} = \frac{ \pi^{\frac{d-1}{2}} \cdot {\zeta(d+1)} \cdot \Gamma(d+2)}{\Gamma(\frac{d+1}{2})} \cdot \frac{g k_{\rm b}^{d+1} T^{d}}{h^d v^{d-1}}.}
    \label{generalized_kapitza}
\end{eqnarray}
Since $\zeta(s) = \sum_{u=1}^{\infty} 1/u^s$, we have $\zeta(2)=\pi^2/6$ for $d=1$,  $\zeta(3) \approx 1.202$ for $d=2$, and $\zeta(4)=\pi^4/90$ for $d=3$.  The resulting expressions for the quantum limit of the thermal boundary conductance are therefore $\kappa_{\rm c}^{\rm q} ({\rm 1D}) = \frac{\pi^2 k_{\rm b}^2}{3 h}T$, $\kappa_{\rm c}^{\rm q} ({\rm 2D}) = \frac{12 \zeta(2) k_{\rm b}^3}{h^2 v}T^2$, and $\kappa_{\rm c}^{\rm q} ({\rm 3D}) = \frac{4\pi^5 k_{\rm b}^4}{15 h^3 v^2}T^3$.  Equation \ref{generalized_kapitza} is thus the generalized Kaptiza conductance for $d-$dimensions.  Example values of this limit of thermal boundary conductances at $T=300$ K for 1D is $\kappa_{\rm c}^{\rm q} ({\rm 1D}) \approx 2.8 \times 10^{-10}$ W/K independent of the acoustic velocity $v$.  For $v=10^{4}$ m/s, $\kappa_{\rm c}^{\rm q} ({\rm 2D}) \approx 0.78 $ W/(m$\cdot$K), and $\kappa_{\rm c}^{\rm q} ({\rm 3D}) \approx 2.7 \times 10^{9} $ W/(m$^2$$\cdot$K).  The thermal conductances for multiple branches, for example transverse and longitudinal, must be added appropriately to obtain the net conductance.  Like the quantum limit of electrical contact resistances, practical thermal boundary resistances will be higher than the lower bounds of Equation \ref{generalized_kapitza} due to phonon reflection at the boundary.  A comparison with experimental values\cite{schwab2000measurement} and practical interfacial effects such as non-unity transmission coefficients\cite{chen2005nanoscale} will be discussed in a separate work.  By replacing the sound velocity with the speed of light ($v \rightarrow c$) in Equation \ref{generalized_kapitza} one obtains the thermal boundary conductance when phonons convert to photons as alternate modes of interfacial thermal conductance (e.g. see \cite{guo2021quantum,hutchins2025interfacial}).

Table \ref{tab:quantum_limits_electrical_contact_resistance_thermal_boundary_resistance} summarizes the results of this work.  As a final point, we note that the combination $\frac{\kappa_{\rm c}^{\rm q}}{\sigma_{\rm c}^{\rm q} \cdot T^{d}}$ of temperature with the quantum limits of electrical contact conductance in the degnerate limit, and thermal boundary resistance, is independent of temperature and depends on fundamental parameters ($q, k_{\rm b}, h$), on electronic bandstructure parameters $(g_{\rm s}, g_{\rm v})$, the carrier density, and the phonon dispersion via the phonon velocity $v$.  For 1D, this combination $\frac{\kappa_{\rm c}^{\rm q}}{\sigma_{\rm c}^{\rm q} \cdot T} = \frac{\pi^2}{3g_{\rm s}g_{\rm v}} \cdot \frac{k_{\rm b}^2}{q^2} $ appears similar to the Wiedemann-Franz law.  This is surprising because the original Wiedemann-Franz law has the thermal conductivity due to electrons (fermions) whereas here it is due to phonons (bosons).

\section*{References}
\bibliography{reference}

\end{document}